\documentclass[graybox]{svmult}
\pdfoutput=1

\usepackage{mathptmx}       
\usepackage{helvet}         
\usepackage{courier}        
\usepackage{type1cm}        
\usepackage{makeidx}         
\usepackage{graphicx}        
\usepackage{multicol}        
\usepackage[bottom]{footmisc}

\usepackage{natbib}         

\makeindex             

\begin{document}

\title*{Generalized flows around neutron stars}
\author{Ayan Bhattacharjee}
\institute{Ayan Bhattacharjee \at S. N. Bose National Centre for Basic Sciences, Block -JD, Sector -3, Salt Lake, Kolkata 700106, India, \email{ayan12@bose.res.in}
}
%
%
\maketitle

\abstract*{In this chapter, we present a brief and non-exhaustive review of the developments of theoretical models for accretion flows around neutron stars. A somewhat chronological summary of crucial observations and modelling of timing and spectral properties are given in sections 2 and 3. In section 4, we argue why and how the Two-Component Advective Flow (TCAF) solution can be applied to the cases of neutron stars when suitable modifications are made for the NSs. We showcase some of our findings from Monte-Carlo and Smoothed Particle Hydrodynamic simulations which further strengthens the points raised in section 4. In summary, we remark on the possibility of future works using TCAF for both weakly magnetic and magnetic Neutron Stars.
}

\abstract{In this chapter, we present a brief and non-exhaustive review of the developments of theoretical models for accretion flows around neutron stars. A somewhat chronological summary of crucial observations and modelling of timing and spectral properties are given in sections 2 and 3. In section 4, we argue why and how the Two-Component Advective Flow (TCAF) solution can be applied to the cases of neutron stars when suitable modifications are made for the NSs. We showcase some of our findings from Monte Carlo and Smoothed Particle Hydrodynamic simulations which further strengthens the points raised in section 4. In summary, we remark on the possibility of future works using TCAF for both weakly magnetic and magnetic Neutron Stars.}

\section{Introduction}
\label{sec:1}
In this chapter, we present a brief and non-exhaustive review of the developments of theoretical models for accretion flows around neutron stars (NSs). A somewhat chronological evolution of crucial observations and modelling of timing and spectral properties are given in sections 2 and 3, respectively. In section 4, we discuss the prospect of the Two-Component Advective Flow (TCAF) solution given by Chakrabarti and Titarchuk in 1995 (hereafter, CT95), around NSs, with certain modifications required for NS. In section 5, we highlight some of our findings from Monte-Carlo simulation of spectra of an accreting NS (Bhattacharjee \& Chakrabarti 2017, hereafter BC17). In section 6, we discuss the recent results  from Smoothed Particle Hydrodynamics simulations in presence of cooling where we studied the time evolution of accretion around NSs. Finally, we summarize the key results to argue that the generalized flows around NSs, are likely to be some form of TCAF.

The modelling of accreting neutron stars were largely prompted by significant observational findings in the last 50 years. Although many of the models are of great phenomenological importance, it is impossible to cover all of those and their modified versions, in this text. We take a separate approach to address the evolution of such theories: we point to the observational evidence that either proved the validity of a theory, prompted a modification or refuted a theory in light of new results. This approach lets us to touch upon the relevant models as well as compare them in terms of predictions of observational results. 

\section{Evolution of theories: Timing Properties}
\label{sec:2}

The temporal variations of accreting NSs (both magnetic and weakly-magnetic) are reflected in the Power Density Spectra (PDS) of the lightcurves at different energies. The presence and evolution of Quasi-Periodic Oscillations (QPOs) reveal significant details about both the hydrodynamic and radiative transfer processes. In order to discuss the chronological developments in timing studies, we briefly define (Wang 2016) different classes of QPOs and their observed properties.

\begin{svgraybox}
\begin{itemize}
\item \textbf{Low-Frequency QPOs (LFQPO):} Low-frequency QPOs are observed in the range of 5 - 60 Hz, with Quality Factor $Q \geq 2$, and amplitudes of $1\%-10\%$. The Horizontal-, Normal-, and Flaring Branch Oscillations all lie in this domain and are respectively abbreviated as HBO, NBO and FBO.
\item \textbf{hecto-Hz QPOs (hHz QPO):} The hecto-hertz (hHz) QPO is usually a peaked noise. If the peak is coherent enough to have a quality factor $Q>2$, it is classified as a QPO. The frequency $\nu_{hHz}$ is seen in the range of $100-200$ Hz. It has rms amplitude between $2\%$ and $20\%$.
\item \textbf{Kilo Hz QPOs (kHz QPO):} These are defined as the QPOs in the range $200~Hz < \nu_{kHz} < 1300~Hz$. 
\end{itemize}
\end{svgraybox}

The first detailed study of a magnetic neutron star, with almost radial accretion, was carried out by Elsner \& Lamb (1977). They studied the formation of magnetosphere, its scale and structure, the stability of its boundary and the plasma flow into the magnetosphere and Alfven surface. Ghosh, Lamb \& Pethick (1977) studied the spin-up and spin-down mechanism of the star due to this. Ghosh \& Lamb (1979a, 1979b) focused on the radial and vertical structures of the transition region and the changing of period of the pulsating stars. Van der Klis et al. (1985) studied the power density spectra of the source GX 5-1, and found intensity dependence of the centroid frequency, width and power of the observed QPOs (20 Hz and 40 Hz) and low-frequency noises ($< 15~Hz$). Lamb et al. (1985) found QPOs in 5-50 Hz range for Sco X-1. They proposed that a clumpy disc, while interacting with a weak magnetosphere, produced the modulation in accretion rate and oscillations in X-ray flux. Priedhorsky et al. (1986) found a bimodal behavior in the response of QPO frequency with intensity. The 6 Hz QPO, in quiescent state, was anti-correlated with intensity. The QPO in the active state, however, was correlated with intensity and varied from 10-20 Hz. Two modes of spectral behaviors were also identified for Sco X-1. Paczynski (1987) suggested the luminosity variations of the boundary might be a result of a unsteady flow with low viscosity. Van der Klis et al. (1990) studied the atoll source 4U 1636-53. Correlation between spectral shape, fast variability, burst duration and temperature with accretion rate $\dot{M}$ hinted that the source state is determined by the accretion rate $\dot{M}$. The persistent intensity varied independent of other characteristics, indicating that intensity was not probably a good measure of accretion rate. Strohmayer et al. (1996) studied the burst and quiescent states of the LMXB 4U 1728-34 to find pulsations of 363 Hz and twin QPOs, respectively. The twin kHz QPOs varied between 650-1100 Hz, with count rate till reaching a maximum, while maintaining a constant separation of 363 Hz. The magnetospheric beat frequency model was used to explain the X-ray variability. Van der Klis et al. (1997) found the twin kHz QPO frequency separation is not constant, rather varied from 310 to 230 Hz, for Sco X-1. It was concluded that any beat frequency model or photon bubble model was unlikely to explain the origin of these peaks. Authors also found HBOs near 45 Hz and 90 Hz, both of which increased with accretion rate. In M{\'e}ndez et al. 1997, the energy dependent study of twin kHz QPO and band limited noise revealed different behaviors, suggesting the two processes are generated at two different radii. It was remarked that the kHz QPOs were due to oscillations of the boundary layer, very close to the surface and the broad noise was generated from near the inner edge of a disc. It was also mentioned that QPO frequency and the noise cut-off frequencies tracked $\dot{M}$ better than the count rate. Wijnands et al. (1997) found the coexistence of HBOs and kHz QPOs for GX 17+2. This suggested that the magnetospheric beat frequency model could not address both the phenomena simultaneously. The rms and FWHM of the upper kHz QPO varied with the change in frequency, whereas the similar quantities remained almost constant for lower kHz QPO with the change of frequency. Wijnands et al. (1998) found both kHz QPOs and HBOs for the Z-source Cyg X-2. Jonker et al. (1998) showed that the twin kHz QPO frequencies moved to higher values with the increase of accretion rate for the Z-source GX 340+0. However, the rms and FWHM of lower QPO remained consistently constant, whereas the ones corresponding to upper kHz QPO decreased. Simultaneous HBO was also detected along with its second harmonic between 20 to 50 Hz and 38 Hz and 69 Hz. It was also concluded from the study of FWHM of HBOs with states, that something other than $\dot{M}$ determined the timing properties. Wijnands et al. (1998) found very similar results for the object GX 5-1, where simultaneous kHz QPOs and HBOs were observed. 

In Titarchuk et al. (1998, hereafter TLM98) a super-Keplerian transition layer was invoked to explain the kHz QPOs (a detailed discussion to be followed in the next section). M{\'e}ndez et al. (1998) showed that the LMXB 4U 1608-522 had twin kHz QPOs and the frequency separation is not constant with the change of centroid frequencies, contradicting a simple beat-frequency interpretation. The authors compared different models for the accretion onto this atoll source having $10\%$ average luminosity of Sco X-1, a Z-source showing similar features, and concluded that the photon-bubble model, the transition layer model nor any modified beat frequency model were able to explain the phenomena. M{\'e}ndez \& van der Klis (1999) found simultaneous existence of burst oscillations and twin kHz QPOs for the source 4U 1728-34. The frequency separation always remained smaller than the burst frequency, for varying accretion rates. For the same object, Titarchuk \& Osherovich (1999) used the transition layer model to identify the low-Lorentzian frequency to be due to radial oscillation in viscous time-scale and the break frequency to be associated with radial diffusion time-scale. Psaltis et al. (1999) studied many non-pulsating NSs and black holes (BHs) of varying luminosities and source types. A correlation between 1)the NS: kHz QPOs and HBOs, 2)BH: QPOs and noises, of same type but varying 3 orders of magnitude in frequency and coherence was found. This suggested that the variations are systematic and related by similar processes in the two types of sources, further constraining the  theoretical models of these phenomena. Titarchuk et al. (1999) added the effects of coriolis forces and interaction of magentosphere with flow to explain the different modes of QPOs observed for 4U 1728-34. Dieters et al. (2000) remarked that the position on the Z-track or the spectral state was not the only parameter that governs the behavior of Sco X-1. Jonker et al. (2000) discovered a new broad component in the PDS of GX 340+0 between 9 Hz to 14 Hz. Yu et al. (2001) studied the source Sco X-1 and found that the upper kHz QPO frequency and the ratio of lower to upper kHz QPO amplitude was anti-correlated with the count rate that varied in the NBO timescale (6-8 Hz). This suggested that some of the NBO flux was generated from inside the inner disk radius and the radiative stress modulated the NBO frequency. Muno et al. (2000) did a longterm study of a total of 13 Z and atoll sources to remark that the atoll sources trace out the color-color diagram over a larger range of luminosity, on a much longer timescale and have a harder spectra when they are faint as compared to the Z-sources. But, both types trace out similar three branched color-color patterns, suggesting a similarity, which was previously not recorded because of incomplete sampling.

Belloni et al. (2002) conducted a detailed analysis of PDS of multiple BHCs and NSs, by decomposing the PDS onto 4 primary Lorentzians. They showed that this not only provides a better statistics, but also yields a better phenomenology which tracked the evolution of different QPO frequencies with spectral states. It was found that a one-on-one correspondence can be found between different QPOs in NSs and BHs, which obey similar correlations. This suggested that, apart from the mass dependence of the dynamical timescale, the physical processes governing the phenomena could be the same and the high-frequency components might not require the NS surface as black holes also show similar features with weaker amplitude. Mauche (2002) drew a similar conclusion by studying the correlations between two QPOs in BHCs, NSs and White Dwarfs (WDs) which showed that the correlated nature can be extended another two orders of magnitude with WDs. The author discussed several models in light of the observed QPOs and dwarf-nova oscillations (DNOs). All models that relied on the presence of strong magnetic fields or a stellar surface to explain the high frequency oscillations were ruled out. The only one which produced the observed frequencies was the TL model where Keplerian orbital frequencies are suggested as the origin of QPOs.

Wijnands et al. (2003) studied multiple accreting neutron stars which had burst oscillations. It was found that any simple beat frequency model or any model which relied on general relativistic effects very close to the rotating star is unlikely to explain the phenomena. The authors stated that the QPOs can be understood in terms of resonances at privileged radii in an accretion disc, but pointed towards general relativistic epicyclic motions to be the cause of such resonance. Barret et al. (2005) studied the Q factor and coherence time variation of both the kHz QPOs to reveal that the Q factor follow different pattern in the two. From their analysis of the LMXB 4U 1608-52, that any model involving clumps orbiting within or above the accretion disk were ruled out, and disc/shock oscillations were suggested as the more likely mechanism. Belloni et al. (2005) showed that the ratio of upper and lower kHz QPO frequencies was found to be 3/2 only because of a biased sampling. A simple resonance model of a Keplerian disc, thus, is unlikely to address the kHz QPOs. Cassella et al. (2006) studied the continuous transition from FBO to NBO in $\sim 100 s$ timescale, for Sco X-1. From the one-to-one correspondence between the LFQPOs in BHs and NSs, it was inferred that the physical mechanisms that determine these oscillations, which are present in both, are in favor of a disk origin of oscillation and rules out models involving interaction with the surface or magnetosphere.

M{\'e}ndez (2006) studied, for multiple NSs, the maximum Q factor and rms power associated with upper and lower kHz QPOs. For the lower one, $Q_{max}$ increased at first and decreased exponentially with increasing luminosity. For the upper one, the Q value remained almost independent of luminosity of the sources. The author suggested that one possible mechanism of generation of kHz QPOs would be that the mass accretion rate sets the size of the inner radius of the disk which determines the QPO frequency as well as the relative contribution of the high-energy part of the spectrum to the total luminosity. Boutloukos et al. (2006) also pointed towards high radial accretion instead of a Keplerian disc, to explain the relatively lower value of kHz QPO frequencies found in the source Cir X-1. Wang et al. (2012) studied the energy dependence of the low-frequency NBOs for Sco X-1. They concluded that in near Eddington luminosities, the oscillations in the TL, the region between inner edge of disk and neutron star, might be responsible for the NBOs. It was noted that the centroid frequency varried non-monotonically with energy indicating a radial oscillation.

\section{Evolution of theories: Spectral Properties}
\label{sec:3}

The explanation of soft state spectra of NSs demanded the presence of a blackbody emission from the boundary layer of a neutron star (Mitsuda et al. 1984). For the harder states with a power-law tail in the energy spectrum, the need of Compton scattering became evident (White et al. 1986, Mitsuda et al. 1989). The difference between these two models was that while the former assumed a cooler boundary layer, the latter assumed a hotter one, compared to the accretion disc. Sunyaev and his collaborators (Inogamov and Sunyaev, 1999; Popham and Sunyaev, 2001; Gilfanov and Sunyaev, 2014) assume that the Keplerian disk reaches all the way to the NS and is connected with the boundary layer where the thickness increases due to higher temperature. Most of these studies were done to address the soft state spectra of neutron stars. The state transition of neutron stars in LMXBs, presented another problem. The fact that disk accretion rate was not the single factor that controlled the size or temperature of the Compton cloud, used to model the hard state spectra, lead to the conclusion that some unknown parameter, related to the truncation radius of the disc, is responsible for the hard X-ray tail (Barret 2001, Barret et al. 2002, Di Salvo and Stella 2002). Paizis et al. (2006) found a systematic positive correlation between the X-ray hard tail and the radio luminosity,  inferring that the Compton cloud might serve as the base of radio jets (which is same as proposed for black holes by Chakrabarti and his group. See Chakrabarti, 2017 and references therein for  a review). Recent phenomenological works places a TL or Compton cloud between the Keplerian disk and the boundary layer (Farinelli et al. 2008; Titarchuk et al. 2014, hereafter TSS14).  It has been argued in the past (Chakrabarti, 1989; Chakrabarti, 1996; Chakrabarti \& Sahu, 1997) that while in black hole accretion, passing of the flow through the inner sonic point ensures that the flow becomes sub-Keplerian just outside the horizon, in the case of NSs, the Keplerian flow velocity must slow down to match with the sub-Keplerian surface velocity. Numerical simulations clearly showed that jumping from a Keplerian disk to a sub-Keplerian disk is mediated by a super-Keplerian region (Chakrabarti \& Molteni, 1995). In TSS14 the TL was expanded several fold to explain the spectral properties.  

Monte-Carlo simulations are essential in generating and understanding spectra emergent from highly non-local processes such as Comptonization. Toy models were made of spherical Compton clouds of constant temperature and optical depth, surrounding a weakly magnetic neutron star to generate hard X-ray tails (Seon et al. 1994).  In case of neutron stars with strong magnetic fields ($B\sim 10^{10-12}~Gauss$), matter lands at the poles through the accretion column and the use of such a geometry leads to the successful explanation of spectral properties (Odaka et al. 2013, Odaka et al. 2014) in certain cases.  Although these studies provide some answers, so far, the spectral fitting carried out were based on phenomenological models which used arbitrarily placed Compton cloud.  Since 1995, the entire community of black holes and neutron stars have started to use cartoon models of the flow inspired by the solution of CT95, both for the explanation of spectral and the explanation of QPOs.
 
An extended TL was used to explain spectra using COMPTT and COMPTB models.  Out of the two COMPTB components used, the one corresponding to Comptonization of NS surface photons, showed a saturation in COMPTB models spectral index (Farinelli and Titarchuk 2011).  Many NS LMXBs are studied using this framework, such as 4U 1728-34 (Seifina et al. 2011), GX 3+1 (Seifina and Titarchuk 2012), GX 339+0 (Seifina et al. 2013), 4U 1820-30 (Titarchuk et al. 2013), Scorpius X-1 (TSS14), 4U 1705-44 (Seifina et al. 2015) etc. Recently, the HMXB 4U 1700-37 has also been examined using the same model (Seifina et al. 2016). 

\section{Two-Component Advective Flows around Neutron Stars}
\label{sec:4}

\begin{figure}
\includegraphics[height=5.0cm,width=12.0cm]{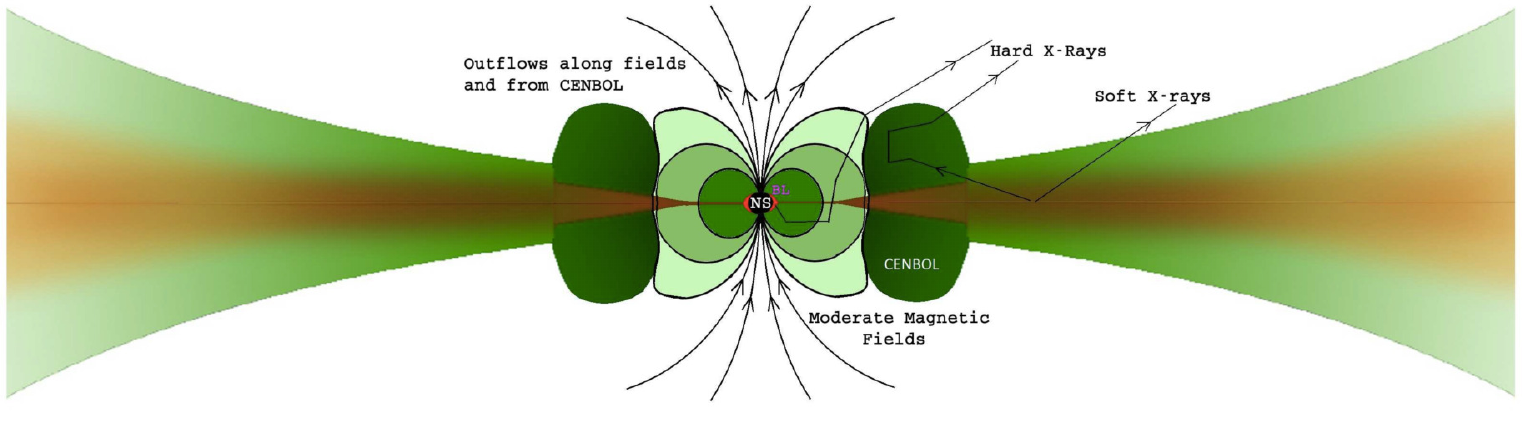}
\caption{A schematic diagram of accretion dynamics and radiation processes in a Two Component Advective Flow solution with a centrifugal barrier supported shock front close to a neutron star with moderate magnetic field. The Normal Boundary Layer (NBOL) on the star surface (BL, in the figure) as a result of a shock, and a secondary Compton cloud in the magnetosphere apart from the possible presence of CENBOL makes the spectra complex (adopted from Chakrabarti 2017).}
\end{figure}

It is well known that the spectral and timing properties of a black hole cannot be explained by a standard Keplerian disk alone (Sunyaev \& Truemper, 1979, Sunyaev and Titarchuk 1980, 1985; Haardt and Maraschi 1993, Zdziarski 2003; Chakrabarti \& Wiita 1993; CT95; Chakrabarti, 1997).  A spectrum clearly has a thermal component resembling of a multi-colour blackbody radiation (Shakura and Sunyaev 1973).  However, the other component is similar to a power-law component which is produced by inverse Comptonization of the thermal or non-thermal electrons (Sunyaev \& Titarchuk, 1980, 1985). There are many models in the literature which present possible scenarios of how the electron cloud might be produced.  However, there is a unique self-consistent solution, namely, the transonic flow solution based TCAF which addresses all the aspects of spectral and temporal properties at the same time. In this scenario, a centrifugal barrier supported boundary layer or CENBOL is produced very close to a compact object in the low viscosity advection component.  The boundary of CENBOL is a shock transition which may or may not be stationary.  The post-shock region is a natural reservoir of hot electrons. Higher viscosity flow component near the equatorial plane becomes a Keplerian disk (Giri \& Chakrabarti 2013) and emits a multi-colour blackbody radiation which are intercepted and re-radiated by the CENBOL to create the power-law like component (Garain et al. 2014) which normally has an exponential cut-off where recoil becomes important. 

When the shock oscillates due to resonance (Molteni et al. 1996, hereafter MSC96) or non-satisfaction of Rankine-Hugoniot condition (Ryu et al. 1997), the resulting hard X-ray intensity is modulated with the size of the CENBOL and manifest as the low-frequency quasi-Periodic oscillations (LFQPOs). The CENBOL is also the source of outflows and jets, in that, when the CENBOL is cooled down due to excessive soft photons, the jet itself also disappears. There are clear observational evidence of two component advective flows in several black hole candidates (Smith et al. 2001; Smith et al. 2002; Debnath et al. 2013; Mondal et al. 2014; Dutta \& Chakrabarti, 2016; Bhattacharjee et al. 2017).  

With an independent advective flow as in TCAF, the phenomenological Compton cloud is naturally explained without having to modify the earlier models drastically. Indeed, since the flow at the outer edge of the disk has little knowledge of the nature of the compact source, except near the innermost boundary, the overall flow configuration is not expected to be very different, especially when the magnetic field is weak ($<10^8$~Gauss). The gravity simply lets the advective matter to fall almost freely till the surface of the star is hit. In reality, there are two such layers simultaneously present in a neutron star accretion (see Fig. 1): One is similar to the normal boundary layer (NBOL) and the other is similar to the CENBOL in a black hole accretion (CT95). In a black hole accretion, only CENBOL is present. The similarities between the flow configuration demanded by observation and the TCAF scenario strengthens the conclusion of Chakrabarti \& Sahu (1997) that the solutions of the transonic flows are modified only in the last few Schwarzschild radii as per the boundary condition of the gravitating object.

\begin{svgraybox}
\vspace{-0.5cm}
\subsection{Prospect of TCAF as the generalized flow around NS}
We list some properties of a `generalized' flow around NS which are present in the TCAF:
\begin{itemize}
\item $\nu_{LF}$ bimodal behaviour: Two separate accretion rates control the intensity. One reduces the oscillation radius, another increases it.
\item Correlated QPOs seen in BHs, NSs and WDs: The generalized model should not require the presence of a stellar surface or magnetosphere.
\item Similar $\nu_{kHz}$ in sources of varying intensity: QPOs are controlled by more than one accretion rates.
\item Long coherence time of kHz QPOs: Clumpy disk models with azimuthal asymmetry to generate QPOs in Keplerian orbital time-scale are unlikely. Radial and vertical oscillation would be preferred.
\item A `mass' accretion rate which controls both the inner edge of the disk that generates kHz QPOs also decides the relative contribution to high energy part of the spectrum.
\item The Compton cloud acts as the base of the jet.
\item The state transition was not controlled by a single accretion rate.
\item Low $\nu_{kHz}$ in some NSs: Radial accretion (advective flow) is preferred over a Keplerian disc.
\end{itemize}
\end{svgraybox}

\section{Spectral Modelling}
\label{sec:5}
We explored the spectral properties of TCAF around an NS in BC17. We had computed the effects of thermal Comptonization of soft photons emitted from a Keplerian disk and the boundary layer of the NS by the post-shock region of a sub-Keplerian flow (Ghosh et al. 2009; Chakrabarti 1985), formed due to the centrifugal barrier. In the simulations, the shock location $X_s$ was also the inner edge of the Keplerian disc. A series of realistic spectra were computed after taking care of the photoelectric abosption by the interstellar medium. We assigned a set of electron temperatures of the post-shock region $T_{CE}$, the temperature of the normal boundary layer (NBOL) $T_{NS}$ of the neutron star and the shock location $X_s$. These parameters depend on the cooling due to intercepted soft photons by the CENBOL region and thus controlled by the flux (or temperature) of the two black body emitting regions: NBOL and the disk. The disk and halo accretion rates ($\dot{m_d}$ and $\dot{m_h}$, respectively) controlled the resultant spectra. The computed spectra were similar to those of weakly magnetic neutron stars in hard states (Fig. 2, left panel). We also studied the case of cooling of the CENBOL when $T_{CE}$ value is high ($250~keV$) and it is shown in Fig. 2 (right panel). 

\begin{figure}
\includegraphics[height=6.0cm,width=6.0cm]{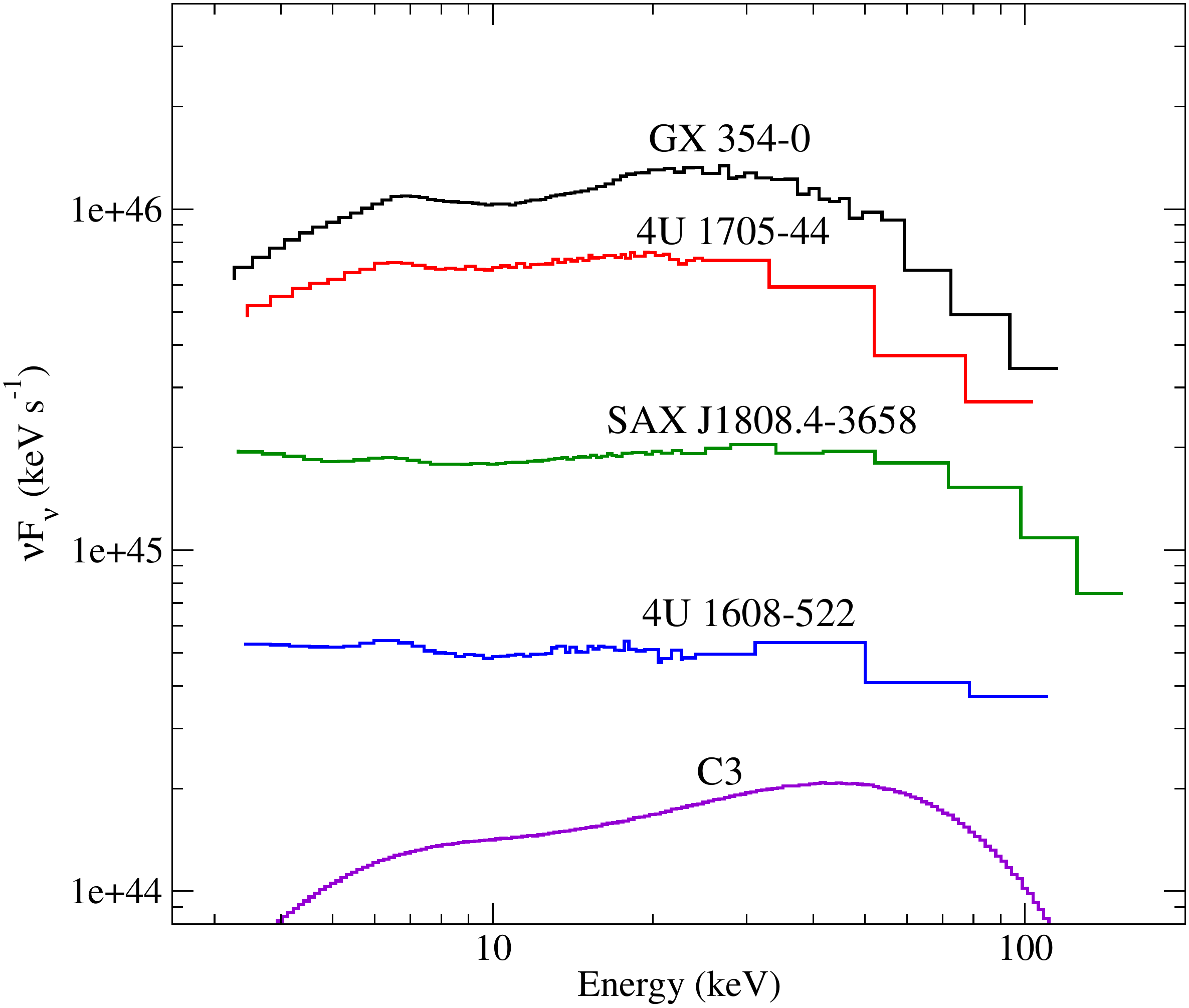}
\includegraphics[width=5.0cm]{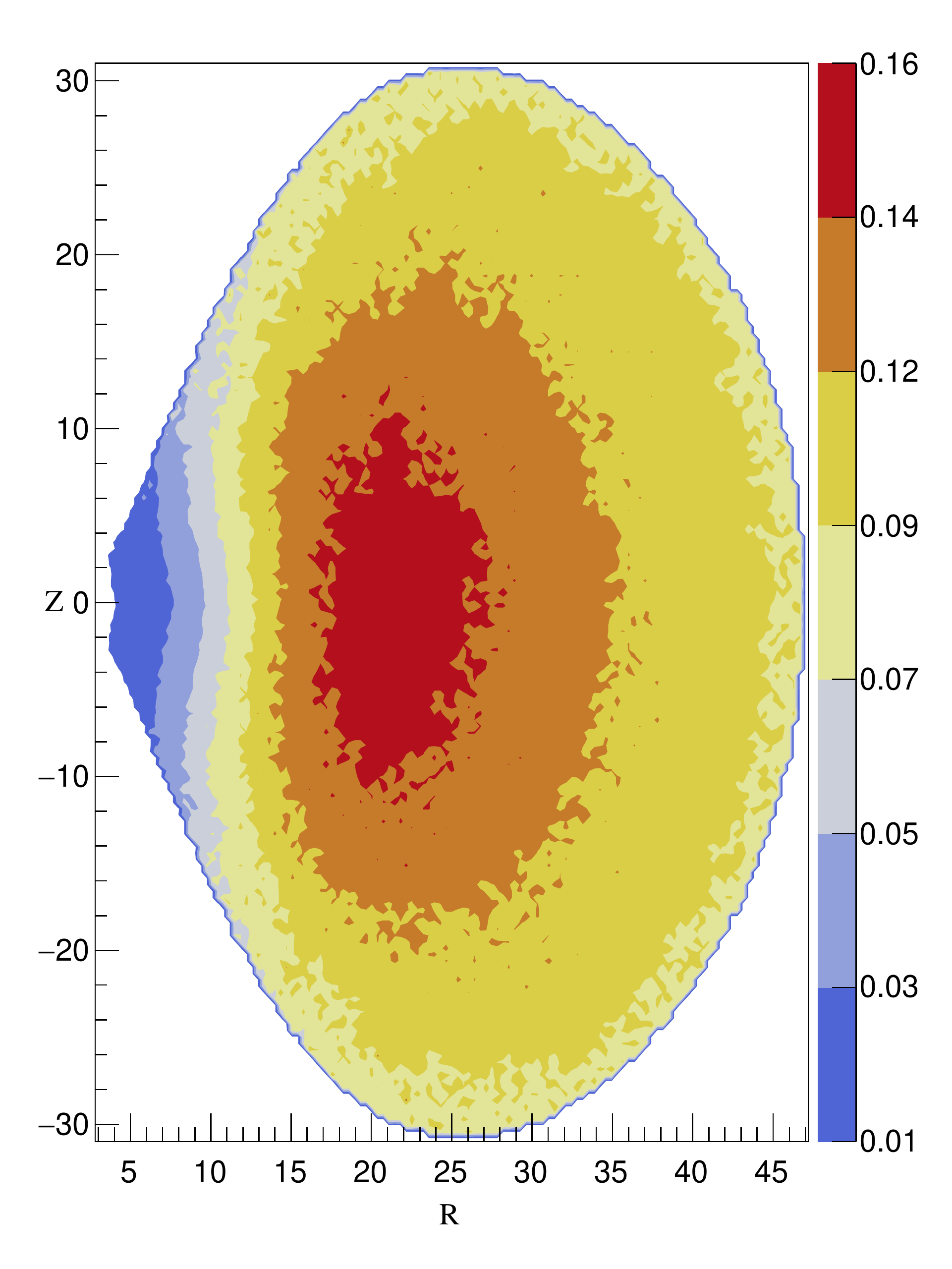}
\caption{\textbf{Left panel:} The spectra of a few weakly magnetized neutron stars. For comparison, we put the spectrum of Case C3 ($\dot{m}_d=0.1,~\dot{m}_h=0.5,~X_s=46.8~r_S,T_{CE}=25.0~keV$) of our simulation. It can be seen that the hard state spectral features are generally reproduced by our simulations. These plotted spectra of different objects were obtained by RXTE observations and are adapted from Gilfanov (2010).
\textbf{Right panel:} The temperature contours after Compton cooling, from the simulation (for initial $T_{CE}=250.0~keV$, $X_s=46.8~r_S$). The temperatures are written in dimensionless unit $(kT_e/mc^2)$.  The halo accretion rate is $0.7$. We observed that with the increase of $\dot{m}_h$, the temperature of CENBOL closer to NBOL, $T_e(\tau_0/2)^{NS}$ decreased. Here $\tau_0$ is the total optical depth of CENBOL calculated along the equatorial plane. The temperature of CENBOL closer to disc, $T_e(\tau_0/2)^{KD}$, also decreased, but remained greater than the corresponding $T_e(\tau_0/2)^{NS}$. The modified contours show the effective region of Compton scattering changes due to higher temperature of CENBOL. The contour in red shows the most effective region for Compton scattering and it is similar to the proposed geometry of TSS14. (Adopted from BC17)}
\label{fig2}
\end{figure}

\textbf{
\begin{svgraybox}
\textbf{Key Findings (BC17):}
\begin{enumerate}
\item The spectrum became harder when $\dot{m}_h$ was increased.
\item We showed the evidence of spectral hardening as the inclination angle of the disk is increased. 
\item The absorption due to ISM was used to convolve the computed spectrum, in $0.1$ to $250$ keV range. We found the spectra to be very similar to realistic hard-state spectra of accreting neutron stars (see Fig. 2). 
\item An increase in accretion rates cools the CENBOL down. It also creates an outward shift of the hottest region. A cooler cloud near NBOL and a hotter cloud near the disc, having equal optical depth$\tau_0/2$ (see Fig. 2), is created. 
\end{enumerate}
\end{svgraybox}
}

\section{Hydrodynamics}
\label{sec:6}
We have used the Smoothed Particle Hydrodynamics (SPH) method (MSC96) to simulate the behavior of inviscid flow accreting on an NS. During the hard spectral state, the incoming flow has a negligible viscosity, leading to the absence of any disk blackbody emission. In such a state, the flow can be assumed to be purely inviscid for the computational purpose. We have shown that for a purely inviscid advective flow having angular momentum, the solution allows two shocks in the flow. The outer one, forms due to strong centrifugal barrier (CENBOL), which is a common feature for both neutron stars and black holes. The inner shock (NBOL) forms very close to the surface of a neutron star, due to the presence of the hard physical boundary. The oscillations in TCAF solution can be of different types depending on the competing forces in play. We have found that an advective flow in presence of cooling can generate many modes of such oscillations in 2D and the numerical values tend to agree with observational results (see Fig. 3, 4). It is to be noted that the infall time scale from CT95 (and subsequent work in the TCAF paradigm) has the same $r^{3/2}$ dependence with the radial distance $r$ and differ from Keplerian orbital time scale by a factor of $R_{comp}/2\pi$, where $R_{comp}$ is the shock strength. This suggests that the TCAF scenario, in principle, would produce the same numerical values of frequencies for all such sources, even when the viscosity is low to form a TL and flow is not-Keplerian to begin with. The pre-shock and post-shock regions of NBOL are orders of magnitude denser than CENBOLs near its outer edge and contribute to oscillations. In the TCAF scenario, the presence of NS surface aids the shock formation but a similar oscillatory features near the compact object are seen in BHs as well since an artificial surface is produced due to the
centrifugal barrier.

\begin{svgraybox}
\textbf{Key Findings:}
\begin{enumerate}
\item The formation of the two shocks in 2D.
\item The formation and variation of outflow rate from the post-shock region when the angular momentum of the inflow varied.
\item Shock formation in strong winds.
\item The radial oscillation of shocks when cooling effects are added.
\item Formation of asymmetries w.r.t. the Z=0 plane and instabilities due to the interaction of inflow and outgoing strong winds which leads to vertical oscillations.
\item The oscillation of NBOL and the variation of effective surface temperature when accretion rate is varied.
\item Presence of the LFQPOs, hecto-Hz QPO and twin kHz QPOs in the power density spectra of Bremsstrahlung loss.
\end{enumerate}
\end{svgraybox}

\begin{figure}
\includegraphics[height=7.0cm,width=3.75cm]{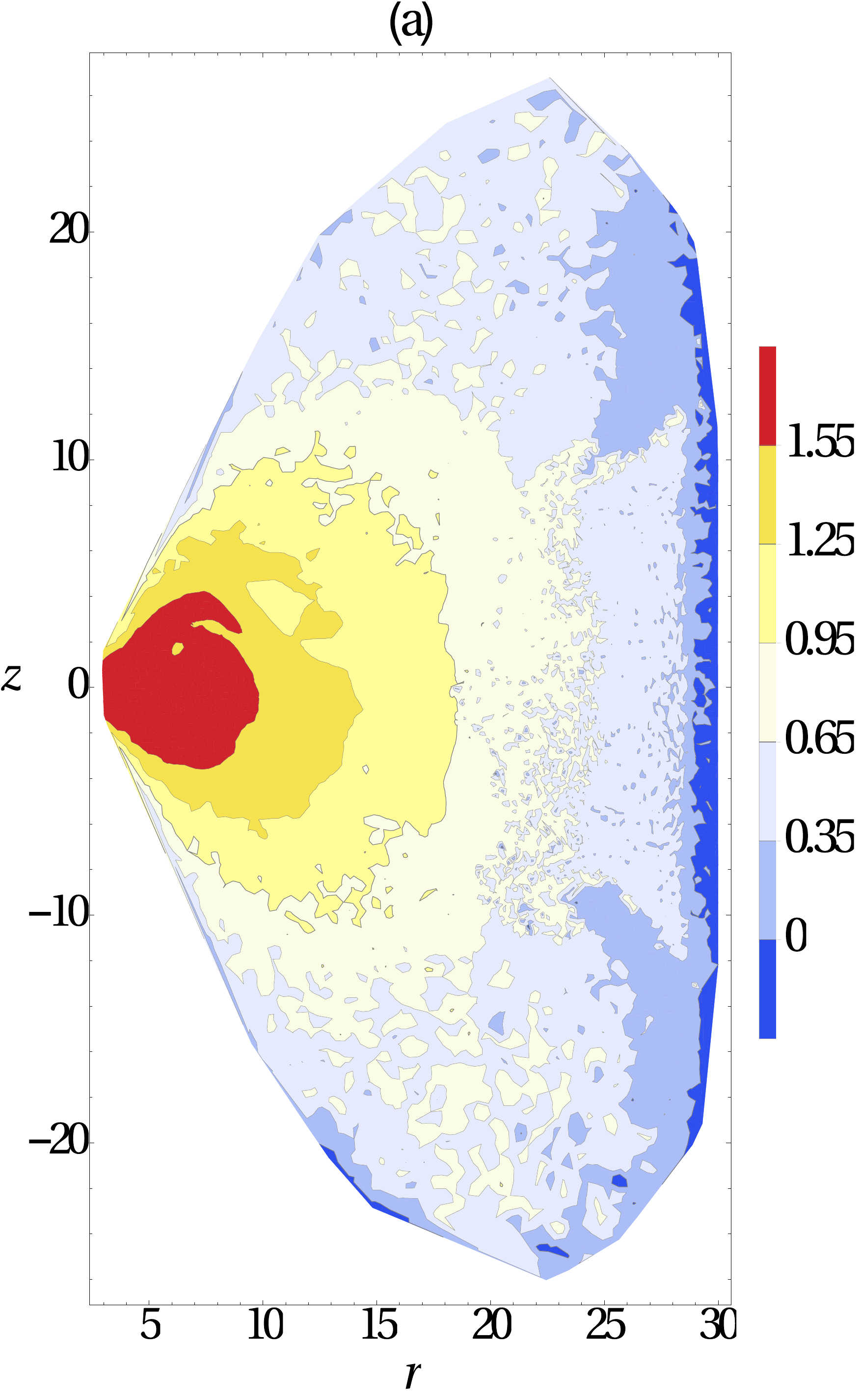}
\includegraphics[height=7.0cm,width=3.75cm]{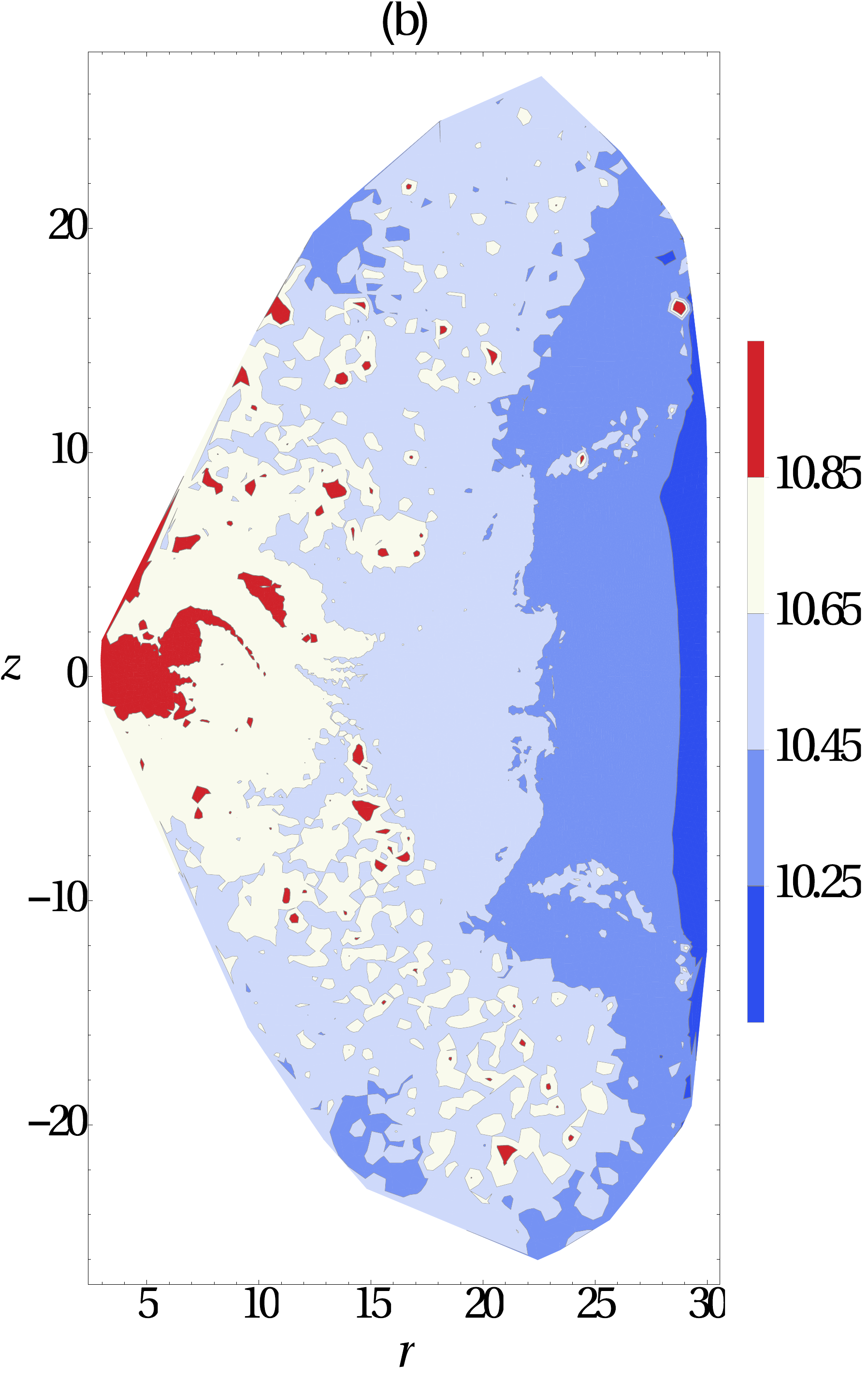}
\includegraphics[height=7.0cm,width=3.75cm]{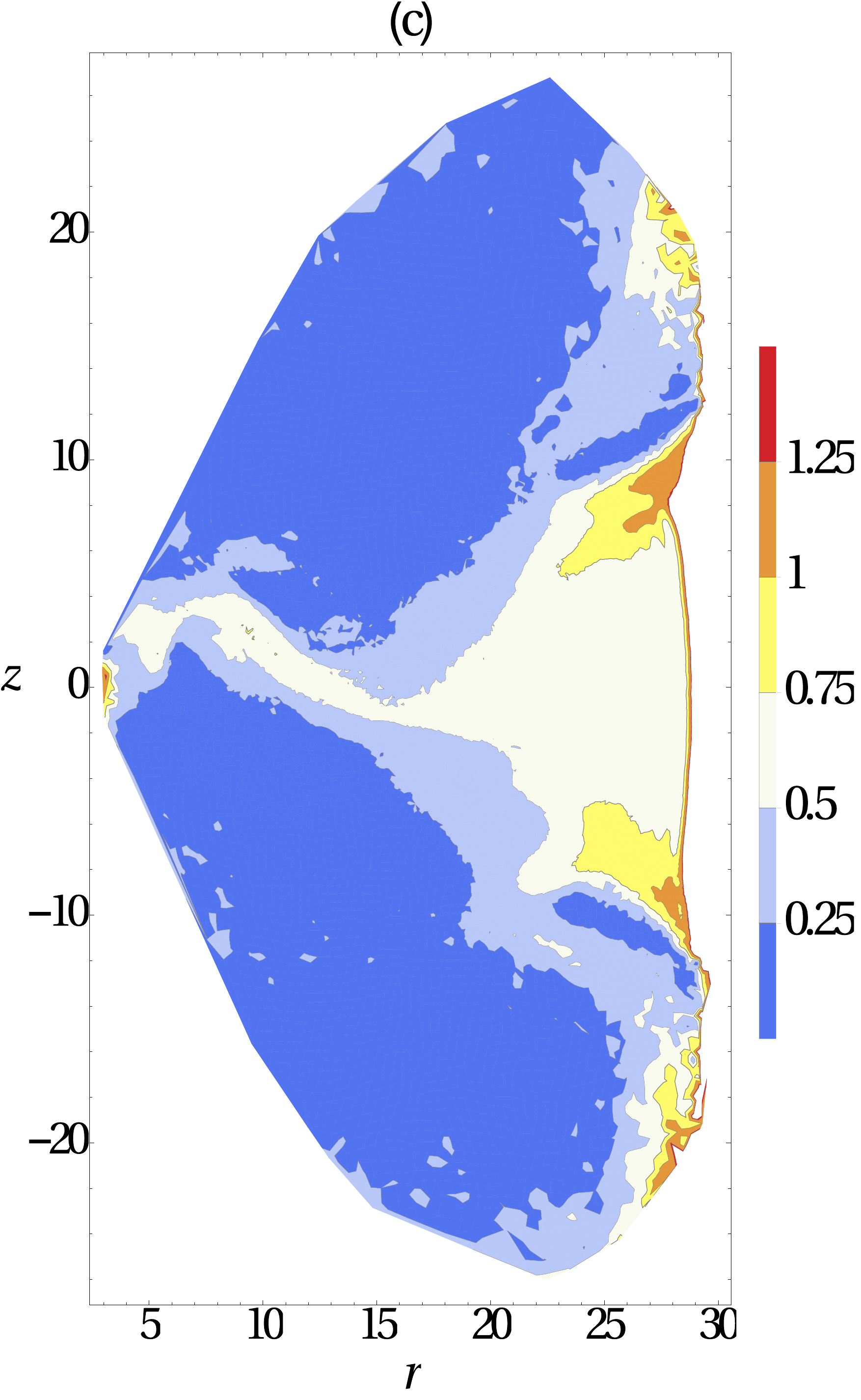}
\caption{From left to right: the density (log scale), temperature (log scale) and Mach number contours for accreting inviscid flow when specific angular momentum $\lambda=1.8$, at $t=0.269 s$ (adopted from Bhattacharjee and Chakrabarti 2018, hereafter BC18). Notice that the Mach number contours near the edge of the star are supersonic and no subsonic contours are plotted. This is due to the fact that in the plotted case, the shock surface was very close to the boundary and the subsonic pseudo particles were absorbed at the surface, before being written out, within the code. The situation, however, is different when viscosity is added, which makes the shock more diffused and a clear supersonic to subsonic transition, comprising of multiple layers of particles are seen. The density and temperature contours are very similar to the thick disk approximation made in BC17, apart from the presence of outflows.}
\end{figure}

\begin{figure}
\sidecaption
\includegraphics[height=7.0cm,width=7.0cm]{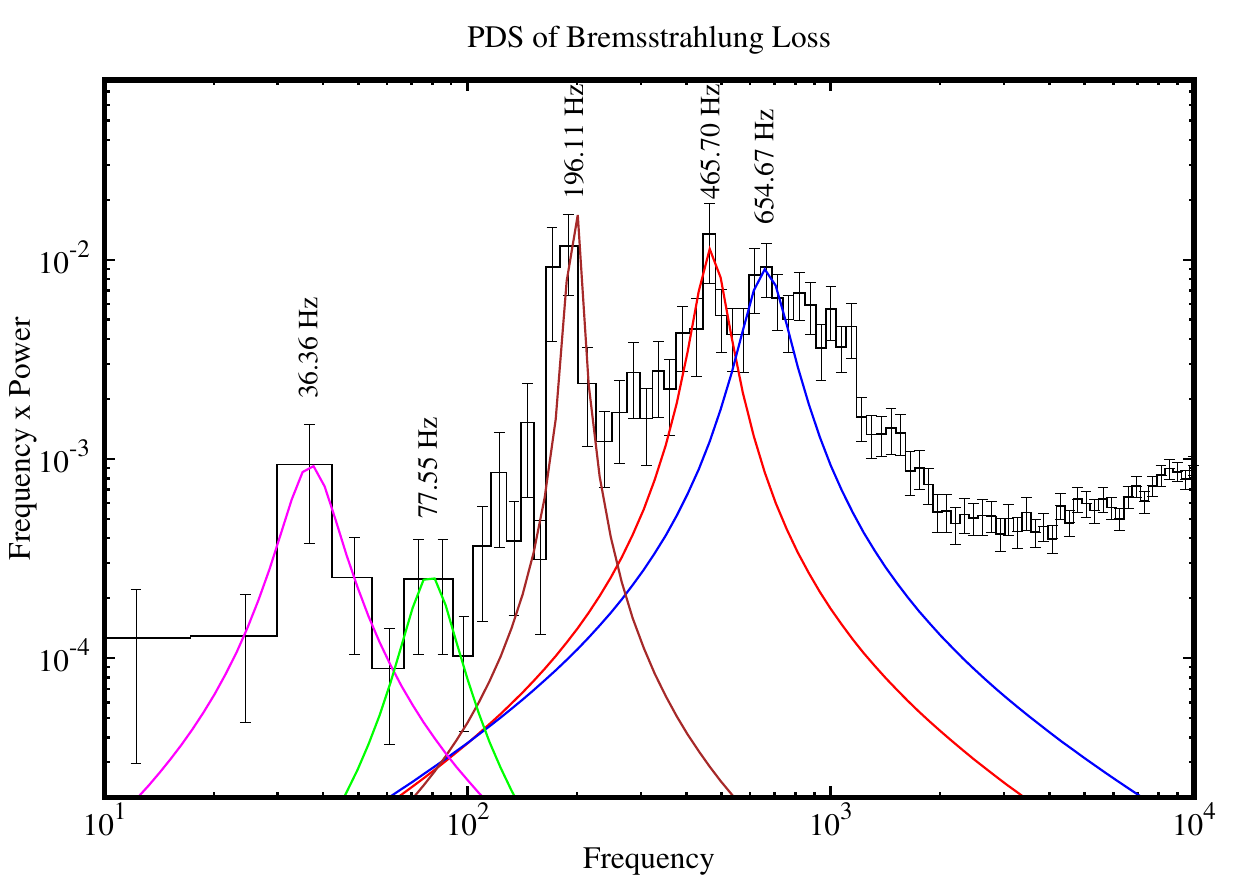}
\caption{The ($frequency \times Power$) vs frequency plot of the Bremsstrahlung loss, is plotted here (adopted from BC18). Five distinct peaks have been fitted with Lorentzian profiles, having centroid frequencies 36.36 Hz (magenta), 77.55 Hz (green), 196.11 Hz (brown), 465.70 Hz (red), and 654.67 Hz (blue) are also plotted. For this simulation $\lambda=1.7,~R_{ns}=4.0~R_s$. It is to be noted that even without invoking any magnetic field and strong viscosity to create a Keplerian disc, our simulations capture the LFQPOs, the hecto-Hz QPO and the twin kHz QPOs.}
\end{figure}

\section*{Concluding Remarks}

The discussions of section 4.1, 5 and 6 prove that the TCAF is at the core of the most generalized flow structure around NS. Since the nature of the companion remains generally similar irrespective of whether the compact object is a BH or an NS, it is natural that we use the TCAF solution for the NSs as well, specifically when the magnetic field is not very strong. As suggested in Chakrabarti 2017 (see Fig. 1), the presence of strong magnetic field will modify the flow configuration from standard TCAF and matter would eventually accrete along the field lines onto the poles, in extreme cases. But, the two components, Keplerian and sub-Keplerian, of accretion would still extend to the magnetosphere, making the flow more complex than what is typically found in the literature so far.

The results of BC17 and BC18 were thought of long ago by Chakrabarti in 1996, where a grand unified solution for advection onto compact objects was discussed. In his own words, if one were to summarize his contributions within a couple of words, it would be realizing the importance of \textbf{`angular momentum'} in accretion. There is no \textit{a priori} reason to assume that companion would eject matter in a fine-tuned Keplerian distribution. The assumption that the flow would have sufficient viscosity which is required to get rid of angular momentum before it reaches a compact object, is also not right. The presence of strong centrifugal force forms shocks and instabilities in the flow without the requirement of extra parameters such as magnetic fields. The sub-Keplerian component of accretion and the effects of cooling have not yet been rigorously implemented in the models of accretion flow around NS. We wish to explore further on this in future.

\label{sec:7}

\begin{acknowledgement}
I would like to acknowledge the constant support and motivation provided by Prof. Sandip K. Chakrabarti. The work we have been doing for the past two years on the numerical simulation of spectral and timing properties of accreting neutron stars, was inspired by his work in the domain of stellar mass and supermassive black holes; conceived by him for the cases of neutron stars due to his unparalleled physical insight into the subject and flexibility to adapt in the face of new observations; implemented by his steadfast approach in supervising our work. I would personally like to thank him for tolerating and answering any and all of my academic questions, irrespective of time or place, for the past 4 years, I have worked under his supervision.
\end{acknowledgement}


\end{document}